# Spin soliton of Holstein model with spin-orbit coupling in one-dimensional conjugated polymers

Shijie Xie    Xiaohui Liu    Qiuxia Lu    and    Sun Yin

School of Physics, State Key Laboratory of Crystal Materials, Shandong University, Jinan 250100, China

## ABSTRACT

For Holstein model with Rashba spin-orbit coupling (SOC) we establish the nonlinear Schrödinger equations and obtain exact soliton solution analytically. It is found that the soliton is spin polarized determined both by the SOC and the electron-phonon (e-ph) interaction. The soliton can be used to describe the spin transport or spin current in organic semiconductors.

## I. INTRODUCTION

Since 2002, organic materials show their application in spintronics, such as organic spin valve [1], organic magnetic field effect [2], organic excited ferromagnetic[3], organic spin selectivity effect [4] and organic spin current [5]. Different from the normal inorganic semiconductors, organic polymers have strong e-ph interactions, which result in the forming of special carriers called solitons and polarons. A soliton or polaron is a self-trapping state of π-electron with lattice, which is spatial localized. It is believed that the spin transport is undertook by the spin solitons or polarons [6], which is much different from that in metals or inorganic semiconductors, where the spin carriers are extended electrons or holes.

A soliton or polaron has been well described in a conjugated polymer with one-dimensional model including the strong e-ph interaction, such as, the tight-binding SSH model [7], PPP model [8] and the continuum TLM model [9]. However, all these models consider the nuclei classically, which may be a non-strict approach as most organic polymers consist of light atoms, such as C,H,O,N. If we consider the quantum effect of the atoms, it will result in an uncertainty of the site position of about $\Delta u \sim \hbar / \Delta p \sim \hbar / \sqrt{2M\varepsilon} \sim 0.005$ nm, which is even larger than the classical value of the dimerization obtained from SSH model [10]. The first all-quantum model including the e-ph interaction is given by Holstein [11]. Although the Hamiltonian is simple, its solving is still full of challenges. In the one-dimensional case, an exact analytical solution is luckily obtained, which describes a spatially localized soliton. During the past years, both the quantized SSH model as well as Holstein model and their interplay have been well studied [12-14].

With the development of organic spintronics, it has been recognized that the spin-orbit coupling (SOC) plays some important roles in organic small molecules as well as conjugated polymers. Usually it is considered that the SOC is weak in organic materials as they consist of light atoms. However, it is also indicated that SOC can be tuned by molecule structure, heavy-metal doping or gate-voltage. For example, the singlet fraction is below 25% in the case of only organic OEP being added as dopant. Nevertheless, it becomes as high as 84% when organometallic compound PtOEP is added as a dopant to the emissive Alq3 film [15]. In addition, in comparison with a device where FIrpic is separated from the DCM2 by BCP, an additional EL is observed when the organometallic FIrpic is used as an additional layer next to the red emitter DCM2 [16]. Yu calculated SOC of some organic molecules and indicated that the SOC can be influenced by how the material is prepared [17]. Therefore understanding the spin property of a soliton or polaron with SOC is critical to realize the spin current and spin manipulation in organic semiconductors. In this paper, we try to give some exact solutions for the Holstein+SOC model. In sec.II we define the model and establish the nonlinear Schrödinger equation with variable principle through trial wave functions. Then the spin soliton solutions are obtained. Finally in Sec.III we conclude.

## II. MODEL AND SOLUTION

Let us consider a doped electron locally coupled to phonon degrees of freedom and employ the Holstein model including the Rashba SOC to describe a one-dimensional polymer chain. The Hamiltonian is written as,

$$H = -t_0 \sum_{i,\sigma} (c^+_{i+1,\sigma} c_{i,\sigma} + c^+_{i,\sigma} c_{i+1,\sigma}) + g \sum_{i,\sigma} c^+_{i,\sigma} c_{i,\sigma} (b^+_i + b_i) + \omega \sum_i b^+_i b_i$$
$$-t_{so} \sum_i \left( c^+_{i+1,\uparrow} c_{i,\downarrow} - c^+_{i+1,\downarrow} c_{i,\uparrow} + h.c. \right) \quad (1)$$

where $c^+_{i,\sigma}(c_{i,\sigma})$ means creation (annihilation) operator of the electron at site $i$ with spin $\sigma$, and $b^+_i(b_i)$ of phonon at site $i$. The first term is the electron hopping between nearest neighbor with hopping integral $t_0$, the second term is the e-ph interaction, the third is the phonon energy with frequency $\omega$, and the fourth term is the SOC. All the parameters have a unit of energy. As the SOC is included, the electronic state will become spin mixed. We write the testing wavefunction in adiabatic approximation as,

$$|\varphi\rangle = \sum_{i,\sigma} \psi_{i,\sigma} c^+_{i,\sigma} e^{\sum_l \left( \alpha_l b_l - \frac{1}{2}\alpha_l^2 \right)} |0\rangle \quad (2)$$

where $|0\rangle$ means the vacuum state. $\psi_{i,\sigma}$ and $\alpha_l$ are two real coefficients. Particularly $\psi_{i,\sigma}$ indicates the probability amplitude of the electron on site $i$ with spin $\sigma$.

With variation method we obtain the equations determining the coefficients as following (see Appendix),

$$\psi_{i+1,\sigma} + \psi_{i-1,\sigma} - D\sigma \left( \psi_{i+1,\bar{\sigma}} - \psi_{i-1,\bar{\sigma}} \right) + 2B \left( \psi^2_{i+1,\uparrow} + \psi^2_{i+1,\downarrow} \right) \psi_{i,\sigma} = C\psi_{i,\sigma} \quad (3)$$

with $\alpha_i = -\frac{g}{\omega} \left( \psi^2_{i,\uparrow} + \psi^2_{i,\downarrow} \right)$, $B = g^2/(\omega t_0)$, $D = t_{so}/t_0$ and $\sigma = \pm$. Eq.(3) is a two-component nonlinear Schrodinger equation group. $C$ corresponds to the eigenvalue, which is determined by the boundary condition.

Taking continuous approximation $\psi_i \to \psi(x)$, we obtain from equ.(3),

$$\psi''_\sigma - 2D\sigma\psi'_{\bar{\sigma}} + 2B \left( \psi^2_\sigma + \psi^2_{\bar{\sigma}} \right) \psi_\sigma = C\psi_\sigma \quad (4)$$

If $D=0$, Eq.(3) becomes a general nonlinear Schrodinger equation. Due to the spin-orbit coupling

($D\neq 0$), The electronic state becomes spin mixing with component $\psi_\sigma$.

Generally, Eq.(4) is not easy to solve. Here we provide its two special solutions:

(a) Extended states:

$$\psi_+ = \begin{pmatrix} \psi_\uparrow \\ \psi_\downarrow \end{pmatrix} = \frac{1}{\sqrt{L}} \begin{pmatrix} \sin(kx) \\ \cos(kx) \end{pmatrix} \quad \text{or} \quad \psi_- = \begin{pmatrix} \psi_\uparrow \\ \psi_\downarrow \end{pmatrix} = \frac{1}{\sqrt{L}} \begin{pmatrix} \cos(kx) \\ \sin(kx) \end{pmatrix},$$

where $k$ means the wave-vector of the electron. The eigenvalue is given by $C_\pm = -k^2 \pm 2Dk + 2B/L$. If there is no SOC, $C_\pm = -k^2 + 2B/L$, the two states are degenerate. But if there is SOC, the degeneracy will be broken with a gap of $4Dk$. It interest to note that, in this case, there is no CDW, i.e., $\rho(x) = |\psi_\pm|^2 = |\psi_\uparrow|^2 + |\psi_\downarrow|^2 = 1/L$. But there is a SDW, i.e., $S(x) = |\psi_\uparrow|^2 - |\psi_\downarrow|^2 = \pm(1/L)\cos 2kx$ with period of $\pi/k$.

(b) Soliton states:

$$\psi_+ = \begin{pmatrix} \psi_\uparrow \\ \psi_\downarrow \end{pmatrix} = \frac{\sqrt{B}}{2} \begin{pmatrix} \sin(Dx)\text{sech}(Bx/2) \\ \cos(Dx)\text{sech}(Bx/2) \end{pmatrix} \tag{5a}$$

or

$$\psi_- = \begin{pmatrix} \psi_\uparrow \\ \psi_\downarrow \end{pmatrix} = \frac{\sqrt{B}}{2} \begin{pmatrix} \cos(Dx)\text{sech}(Bx/2) \\ \sin(Dx)\text{sech}(Bx/2) \end{pmatrix} \tag{5b}$$

These two solutions are degenerate with the same eigenvalue $C_\pm = D^2 + B^2/4$. It is interesting to note that, if the electron is in extended states, the spin degeneracy is broken by SOC. But if the electron is in the localized states, it is found that the spin degeneracy is kept even there is SOC. In this case, the lattice displacement is given by,

$$u(x) = \sqrt{\frac{\hbar}{2M\omega}} \langle \phi | b(x) + b^+(x) | \phi \rangle = -\sqrt{\frac{\hbar}{2M\omega}} \frac{gB}{4\omega} \text{sech}^2 \frac{B}{2} x \tag{6}$$

It is a soliton. The lattice configuration is independent of the SOC. As shown in Fig.1, The localization of the soliton is mainly determined by the e-ph interaction parameter $B$.

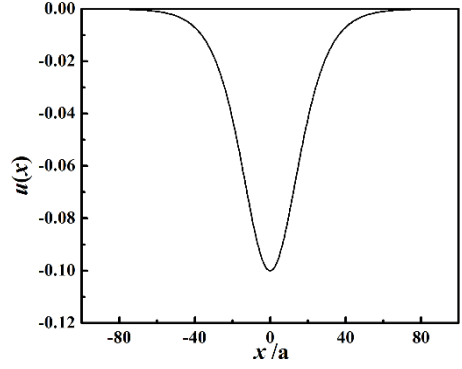

Fig.1 Configuration of the soliton with $B=0.05$.

Spin polarization of the soliton is given by,

$$S(x) = \frac{\hbar}{2}\left[|\psi_\uparrow|^2 - |\psi_\downarrow|^2\right] = \pm\frac{\hbar}{2}\left[\frac{B}{4}\cos(2Dx)\operatorname{sech}^2(Bx/2)\right] \quad (7)$$

The distribution is shown in Fig.2. It is localized and oscillated with a period of $\pi/D$.

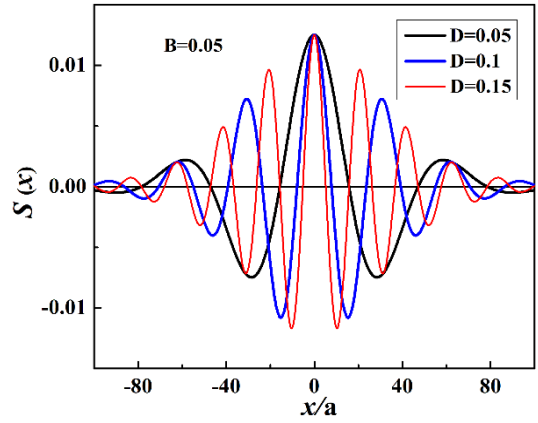

Fig.2 Spin distribution of soliton state.

The total spin polarization of the soliton due to SOC is obtained,

$$S = \pm\frac{\hbar}{2}\left[\frac{B}{4}\int\cos(2Dx)\operatorname{sech}^2(Bx/2)dx\right] \quad (8)$$

If $D=0$, then $S = \pm\hbar/2$, which means that the soliton is in its spin eigenstate. Due to the SOC, the soliton become spin mixing and its spin polarization decreases with increasing SOC. Dependence of the soliton spin upon the SOC is shown in Fig.3. If the e-ph is weak, it is found that the spin polarization vanishes rapidly with the SOC. But if the e-ph is strong, the spin polarization keeps unless the SOC is large enough. Similar results are obtained from the DFT calculation [18] and tight-binding model model [19].

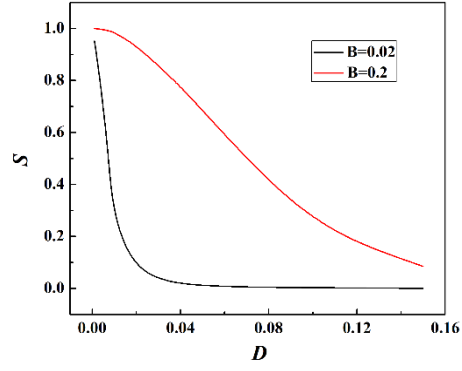

Fig.3 Dependence of the total spin on SOC.

## III. CONCLUSION

In conclusion, Holstein model is one of the simplest models describing a condensed system all-quantitatively. Its solution is a fundamental topic interested by theoretical researchers. Recent years, Holstein model with Rashba spin-orbit coupling in a two-dimensional system has been widely studied with various approaches [20-27]. With the rapid development of organic spintronics these years, it is believed that the spin transport in organic materials is undertook by the soliton through either hopping or exchange mechanism [28,29]. Therefore, exact solution of a soliton in organic polymers is important to understand the spin transport or spin current. Although it is difficult to say whether the Holstein model provides a particularly realistic description of actual organic materials, we investigate its properties with SOC included and partly illustrate the spin features of the soliton exactly.

## APPENDIX: Holstein+SOC model and solutions

For an one-dimensional polymer chain, the Holstein model with Rashba spin-orbit coupling (SOC) is written as,

$$H = -t_0 \sum_{i,\sigma}(c^+_{i+1,\sigma}c_{i,\sigma} + c^+_{i,\sigma}c_{i+1,\sigma}) + g\sum_{i,\sigma} c^+_{i,\sigma}c_{i,\sigma}\left(b^+_i + b_i\right) + \omega\sum_i b^+_i b_i$$
$$-t_{so}\sum_i \left(c^+_{i+1,\uparrow}c_{i,\downarrow} - c^+_{i+1,\downarrow}c_{i,\uparrow} + h.c.\right) \quad (A1)$$

We suggest a testing wavefunction as,

$$|\varphi\rangle = \sum_{i,\sigma} \psi_{i,\sigma} c^+_{i,\sigma} e^{\sum_l \left(\alpha_l b_l - \frac{1}{2}\alpha_l^2\right)}|0\rangle \quad (A2)$$

Where $|0\rangle$ means the vacuum state. $\psi_{i,\sigma}$ and $\alpha_l$ are two real coefficients. Particularly $\psi_{i,\sigma}$ indicates the probability amplitude of π-electrons on site $i$ with spin σ. Hamiltonian (A1) contains four terms. We write their expected values as,

$$E = \langle\varphi|H|\varphi\rangle/\langle\varphi|\varphi\rangle = E_t + E_g + E_\omega + E_{so} \quad (A3)$$

With the normalization,

$$\langle\varphi|\varphi\rangle = \langle 0|e^{\sum_l (\alpha_l b_l - \frac{1}{2}\alpha_l^2)} \sum_{i_1,\sigma_1} \psi_{i_1,\sigma_1} c_{i_1,\sigma_1} \sum_{i_2,\sigma_2} \psi_{i_2,\sigma_2} c^+_{i_2,\sigma_2} e^{\sum_l (\alpha_l b_l^+ - \frac{1}{2}\alpha_l^2)}|0\rangle$$

$$= \langle 0|e^{\sum_l (\alpha_l b_l - \frac{1}{2}\alpha_l^2)} e^{\sum_l (\alpha_l b_l^+ - \frac{1}{2}\alpha_l^2)}|0\rangle \cdot \sum_{i_1,\sigma_1}\sum_{i_2,\sigma_2} \psi_{i_1,\sigma_1}\psi_{i_2,\sigma_2} \delta_{i_1,i_2}\delta_{\sigma_1,\sigma_2} \quad (A4)$$

$$= \sum_{i,\sigma} \psi_{i,\sigma}^2$$

$$E_t = \langle 0|e^{\sum_l (\alpha_l b_l - \frac{1}{2}\alpha_l^2)} \sum_{i_1,\sigma_1}\psi_{i_1,\sigma_1} c_{i_1,\sigma_1}\left[-t_0 \sum_i (c^+_{i,\sigma}c_{i+1,\sigma} + c^+_{i,\sigma}c_{i-1,\sigma})\right]\sum_{i_2,\sigma_2}\psi_{i_2,\sigma_2} c^+_{i_2,\sigma_2} e^{\sum_l (\alpha_l b_l^+ - \frac{1}{2}\alpha_l^2)}|0\rangle$$

$$= -t_0 \langle 0|e^{\sum_l (\alpha_l b_l - \frac{1}{2}\alpha_l^2)} e^{\sum_l (\alpha_l b_l^+ - \frac{1}{2}\alpha_l^2)}|0\rangle$$

$$\times \sum_{i_1,\sigma_1}\sum_{i_2,\sigma_2}\psi_{i_1,\sigma_1}\psi_{i_2,\sigma_2}\langle 0|c_{i_1,\sigma_1}\sum_i (c^+_{i,\sigma}c_{i+1,\sigma}+c^+_{i,\sigma}c_{i-1,\sigma})c^+_{i_2,\sigma_2}|0\rangle$$

$$= -t_0 \sum_{i,\sigma}(\psi_{i,\sigma}\psi_{i+1,\sigma} + \psi_{i,\sigma}\psi_{i-1,\sigma})$$

$$E_g = \langle 0|e^{\sum_l (\alpha_l b_l - \frac{1}{2}\alpha_l^2)} \sum_{i_1,\sigma_1} \psi_{i_1,\sigma_1} c_{i_1,\sigma_1}\left[g\sum_{i,\sigma} c^+_{i,\sigma}c_{i,\sigma}(b_i + b_i^+)\right]\sum_{i_2,\sigma_2}\psi_{i_2,\sigma_2} c^+_{i_2,\sigma_2} e^{\sum_l (\alpha_l b_l^+ - \frac{1}{2}\alpha_l^2)}|0\rangle$$

$$= g\sum_{i,\sigma}\sum_{i_1,\sigma_1}\sum_{i_2,\sigma_2}\psi_{i_1,\sigma_1}\psi_{i_2,\sigma_2}\langle 0|e^{\sum_l (\alpha_l b_l - \frac{1}{2}\alpha_l^2)}(b_i + b_i^+)e^{\sum_l (\alpha_l b_l^+ - \frac{1}{2}\alpha_l^2)}|0\rangle \delta_{i,i_1}\delta_{i_2,i}\delta_{\sigma,\sigma_1}\delta_{\sigma_2,\sigma}$$

$$= 2g\sum_{i,\sigma}\alpha_i\psi_{i,\sigma}^2$$

$$E_\omega = \langle 0|e^{\sum_l (\alpha_l b_l - \frac{1}{2}\alpha_l^2)}\sum_{i_1,\sigma_1}\psi_{i_1,\sigma_1}c_{i_1,\sigma_1}\left[\omega\sum_i b_i^+ b_i\right]\sum_{i_2,\sigma_2}\psi_{i_2,\sigma_2}c^+_{i_2,\sigma_2} e^{\sum_l (\alpha_l b_l^+ - \frac{1}{2}\alpha_l^2)}|0\rangle$$

$$= \omega\sum_i\sum_{i_1,\sigma_1}\sum_{i_2,\sigma_2}\psi_{i_1,\sigma_1}\psi_{i_2,\sigma_2}\langle 0|e^{\sum_l (\alpha_l b_l - \frac{1}{2}\alpha_l^2)} b_i^+ b_i e^{\sum_l (\alpha_l b_l^+ - \frac{1}{2}\alpha_l^2)}|0\rangle\delta_{i_1,i_2}\delta_{\sigma_1,\sigma_2}$$

$$= \omega\sum_{l,\sigma}\psi_{l,\sigma}^2\sum_i(\alpha_i^2 - 1)$$

$$E_{so} = \langle 0|e^{\sum_l (\alpha_l b_l - \frac{1}{2}\alpha_l^2)}\sum_{i_1,\sigma_1}\psi_{i_1,\sigma_1}c_{i_1,\sigma_1}\left[-t_{so}\sum_i(c^+_{i+1,\uparrow}c_{i,\downarrow} - c^+_{i+1,\downarrow}c_{i,\uparrow} + h.c.)\right]\sum_{i_2,\sigma_2}\psi_{i_2,\sigma_2}c^+_{i_2,\sigma_2} e^{\sum_l (\alpha_l b_l^+ - \frac{1}{2}\alpha_l^2)}|0\rangle$$

$$= -t_{so}\sum_{\substack{i,i_1,i_2\\\sigma_1,\sigma_2}}\psi_{i_1,\sigma_1}\psi_{i_2,\sigma_2}\left(\delta_{i_1,i+1}\delta_{\sigma_1,\uparrow}\delta_{i,i_2}\delta_{\sigma_2,\downarrow} - \delta_{i_1,i+1}\delta_{\sigma_1,\downarrow}\delta_{i,i_2}\delta_{\sigma_2,\uparrow} + \delta_{i_1,i}\delta_{\sigma_1,\downarrow}\delta_{i+1,i_2}\delta_{\sigma_2,\uparrow} - \delta_{i_1,i}\delta_{\sigma_1,\uparrow}\delta_{i+1,i_2}\delta_{\sigma_2,\downarrow}\right)$$

$$= -2t_{so}\sum_i\left(\psi_{i+1,\uparrow}\psi_{i,\downarrow} - \psi_{i,\uparrow}\psi_{i+1,\downarrow}\right)$$

Then the energy expected value is,

$$E_{total} = \frac{1}{\sum_{l,\sigma}\psi_{l,\sigma}^2}\left[-t_0\sum_{i,\sigma}(\psi_{i,\sigma}\psi_{i+1,\sigma} + \psi_{i,\sigma}\psi_{i-1,\sigma}) + 2g\sum_{i,\sigma}\alpha_i\psi_{i,\sigma}^2 \right.$$

$$\left. +\omega\sum_{l,\sigma}\psi_{l,\sigma}^2\sum_i(\alpha_i^2 - 1) - 2t_{so}\sum_i\left(\psi_{i+1,\uparrow}\psi_{i,\downarrow} - \psi_{i,\uparrow}\psi_{i+1,\downarrow}\right)\right] \quad (A5)$$

Take variation for $\alpha_k$ and $\psi_{k,\sigma}$ separately, we have,

$$\frac{\partial E_{total}}{\partial \alpha_k} = \frac{1}{\sum_{l,\sigma}\psi_{l,\sigma}^2}\left[2g\sum_{\sigma}\psi_{k,\sigma}^2 + 2\omega\alpha_k\right] = 0$$

which gives,

$$\alpha_k = -\frac{g}{\omega}\sum_{\sigma}\psi_{k,\sigma}^2 \tag{A6}$$

$$\frac{\partial E_{total}}{\partial \psi_{k,\sigma}} = \frac{1}{\sum_{l,s}\psi_{l,s}^2}\left[-t_0\left(\psi_{k+1,\sigma} + \psi_{k-1,\sigma} + \psi_{k-1,\sigma} + \psi_{k+1,\sigma}\right) + 4g\alpha_k\psi_{k,\sigma}\right.$$

$$\left. -2t_{so}\left(\psi_{k-1,\downarrow}\delta_{\sigma\uparrow} + \psi_{k+1,\uparrow}\delta_{\sigma\downarrow} - \psi_{k+1,\downarrow}\delta_{\sigma\uparrow} - \psi_{k-1,\uparrow}\delta_{\sigma\downarrow}\right)\right]$$

$$-\frac{2\psi_{k,\sigma}}{\left[\sum_{l,s}\psi_{l,s}^2\right]^2}\left[-t_0\sum_{i,s}\left(\psi_{i,s}\psi_{i+1,s} + \psi_{i,s}\psi_{i-1,s}\right) + 2g\sum_{i,s}\alpha_i\psi_{i,s}^2\right.$$

$$\left. -2t_{so}\sum_i\left(\psi_{i+1,\uparrow}\psi_{i\downarrow} - \psi_{i+1,\downarrow}\psi_{i,\uparrow}\right)\right]$$

$$= 0$$

Substitute Eq. (A6) into above equation and let $B = g^2/\omega t_0, D = t_{so}/t_0$. By using the normalization of the wavefunction, we obtain,

$$\psi_{k+1,\sigma} + \psi_{k-1,\sigma} + 2B\left(\sum_s\psi_{k,s}^2\right)\psi_{k,\sigma} + D\left[(\psi_{k-1,\downarrow} - \psi_{k+1,\downarrow})\delta_{\sigma\uparrow} + (\psi_{k+1,\uparrow} - \psi_{k-1,\uparrow})\delta_{\sigma,\downarrow}\right]$$

$$-\psi_{k,\sigma}\sum_{i,s}\left(\psi_{i,s}\psi_{i+1,s} + \psi_{i,s}\psi_{i-1,s}\right) - 2B\psi_{k,\sigma}\sum_{i,s}\left(\sum_{s'}\psi_{i,s'}^2\right)\psi_{i,s}^2 - 2D\psi_{k,\sigma}\sum_i\left(\psi_{i+1,\uparrow}\psi_{i\downarrow} - \psi_{i+1,\downarrow}\psi_{i,\uparrow}\right)$$

$$= 0$$

$$\tag{A7}$$

Supposing that,

$$\psi_{k+1,\sigma} + \psi_{k-1,\sigma} + 2B\left(\sum_s\psi_{k,s}^2\right)\psi_{k,\sigma} + D\sigma\left(\psi_{k-1,\bar{\sigma}} - \psi_{k+1,\bar{\sigma}}\right) = C\psi_{k,\sigma} \tag{A8}$$

Substituting Eq.(A8) into Eq.(A7), we will find that, for example, for spin up component,

$$\psi_{k+1,\sigma} + \psi_{k-1,\sigma} + 2B\left(\sum_s \psi_{k,s}^2\right)\psi_{k,\sigma} + D(\psi_{k-1,\downarrow} - \psi_{k+1,\downarrow})$$

$$C\psi_{k,\uparrow} - \psi_{k,\uparrow}\left[\sum_{i,s}\psi_{i,s}\left(\psi_{i+1,s} + \psi_{i-1,s}\right) + 2B\sum_{i,s}\left(\sum_{s'}\psi_{i,s'}^2\right)\psi_{i,s}^2 + 2D\sum_i\left(\psi_{i+1,\uparrow}\psi_{i\downarrow} - \psi_{i+1,\downarrow}\psi_{i,\uparrow}\right)\right]$$

$$= C\psi_{k,\uparrow} - \psi_{k,\uparrow}\left[\sum_i \psi_{i,\uparrow}\left(\psi_{i+1,\uparrow} + \psi_{i-1,\uparrow}\right) + 2B\sum_i\left(\sum_{s'}\psi_{i,s'}^2\right)\psi_{i,\uparrow}^2 + D\sum_i \psi_{i,\uparrow}\left(\psi_{i-1,\downarrow} - \psi_{i+1,\downarrow}\right)\right]$$

$$-\psi_{k,\uparrow}\left[\sum_i \psi_{i,\downarrow}\left(\psi_{i+1,\downarrow} + \psi_{i-1,\downarrow}\right) + 2B\sum_i\left(\sum_{s'}\psi_{i,s'}^2\right)\psi_{i,\downarrow}^2 - D\sum_i \psi_{i\downarrow}\left(\psi_{i-1,\uparrow} - \psi_{i+1,\uparrow}\right)\right]$$

$$= C\psi_{k,\uparrow} - \psi_{k,\uparrow}\left[\sum_i \psi_{i,\uparrow}\left(C\psi_{i,\uparrow}\right) + \psi_{i,\downarrow}\left(C\psi_{i,\downarrow}\right)\right]$$

$$= C\psi_{k,\uparrow} - C\psi_{k,\uparrow}$$

$$\equiv 0$$

Therefore, Eq.(A8) is equivalent to Eq.(A7) and it is the non-linear Schrodinger equation with eigenvalue $C$.

Now let us make continuum treatment for Eq.(A8) by taking,

$$\psi_{k+1} + \psi_{k-1} = (\psi_{k+1} - \psi_k) - (\psi_k - \psi_{k-1}) + 2\psi_k = \psi''(x) + 2\psi(x)$$
$$\psi_{k+1} - \psi_{k-1} = (\psi_{k+1} - \psi_k) + (\psi_k - \psi_{k-1}) = 2\psi'(x) - \psi''(x)$$

And we obtain the continuum form of Eq.(A8) by neglecting the high-order term,

$$\psi_\sigma'' - 2D\sigma\psi_{\bar\sigma}' + 2B\left(\psi_\sigma^2 + \psi_{\bar\sigma}^2\right)\psi_\sigma = C\psi_\sigma \tag{A9}$$

In the continuum approximation, the total energy is

$$E_{total}/t_0 = \sum_\sigma \int (\psi_\sigma')^2 dx - B\sum_{\sigma\sigma'}\int \psi_\sigma^2 \psi_{\sigma'}^2 dx - 2D\int\left(\psi_\uparrow'\psi_\downarrow - \psi_\uparrow \psi_\downarrow'\right)dx \tag{A10}$$

where we omit the constant term $-2 - N\omega$.

**Acknowledgment:** This work was supported by the National Natural Science Foundation of the People's Republic of China (11974211, 11974212) and the Natural Science Foundation of Shandong Province of China (ZR2019MA070).